\documentclass[12pt,preprint]{emulateapj}
\usepackage{amssymb}
\usepackage{graphicx}
\usepackage{epstopdf}

\newcommand{\bma}{\begin{math}}
\newcommand{\ema}{\end{math}}
\newcommand{\beq}{\begin{equation}}
\newcommand{\eeq}{\end{equation}}
\newcommand{\beqa}{\begin{eqnarray}}
\newcommand{\eeqa}{\end{eqnarray}}
\newcommand{\bc}{\begin{center}}
\newcommand{\ec}{\end{center}}
\newcommand{\bit}{\begin{itemize}}
\newcommand{\eit}{\end{itemize}}

\def\n{{\bf  \hat n}}
\def\k{{\bf k}}
\def\l{{\bf l}}
\def\L{{\bf L}}
\def\r{{\bf r}}
\def\x{{\bf x}}
\def\u{{\bf u}}

\def\K{{\rm K}}
\def\mK{{\rm mK}}
\def\arcmin{{\rm arcmin}}
\def\ln{{\rm ln}}
\def\max{{\rm max}}
\def\min{{\rm min}}

\def\iul{{\rm I}}
\def\il{{\tilde{\rm I}}}

\def\pul{{\rm P}}

\def\ptot{{\tilde{\rm P}^{\rm tot}}}
\def\cul{{\rm C}}
\def\cl{{\tilde{\rm C}}}
\def\ctot{{\tilde{\rm C}^{\rm tot}}}

\def\mpch{{\rm Mpc/h}}
\def\mhz{{\rm MHz}}
\def\d2l{\frac{d^2l}{(2\pi)^2}}
\def\dko{\frac{dk_1}{2\pi}}
\def\dkt{\frac{dk_2}{2\pi}}
\def\dtheta{\delta \theta}
\def\dang{\mathcal{D}}
\def\lbox{\mathcal{L}}

\begin{document}

\title{Lensing Reconstruction using redshifted 21 cm Fluctuations}

\author{Oliver Zahn\altaffilmark{1,3}, Matias Zaldarriaga\altaffilmark{1,2}}

\altaffiltext{1} {Harvard-Smithsonian Center for Astrophysics, 60 Garden
  Street, Cambridge, MA 02138, USA}
\altaffiltext{2} {Jefferson Physical Laboratory, Harvard University,
  Cambridge, MA 02138, USA}
\altaffiltext{3} {e-mail address: ozahn@cfa.harvard.edu}

\begin{abstract}

We investigate the potential of second generation measurements of
redshifted 21 cm radiation from before and during the epoch of
reionization (EOR) to 
reconstruct the matter density fluctuations along the line of
sight. To do so we generalize the quadratic methods developed for the
Cosmic Microwave Background (CMB) to 21 cm fluctuations. We show that the three
dimensional signal can be decomposed into a finite number of line of
sight Fourier modes that contribute to the lensing reconstruction. Our
formalism properly takes account of correlations along the line of 
sight and uses all the information contained in quadratic combinations of the signal.
In comparison with the CMB, 21 cm fluctuations have the disadvantage of a 
relatively scale invariant unlensed power spectrum which suppresses the lensing effect. 
The smallness of the lensing effect is compensated by using
information from a range of observed redshifts. 
We estimate the size of experiments that are needed to measure this
effect. With a square kilometer of collecting area  
and a maximal baseline of 3 km one can achieve lensing reconstruction
noise levels an order of magnitude below CMB quadratic estimator
constraints at $L=1000$, and map the deflection field out to less then a
tenth of a degree ($L> 2000$) within a season of observations on one field. 
Statistical lensing power spectrum detections will be possible to
sub-arcminute scales, even with the limited sky coverage that
currently conceived experiments have. One should be able to improve constraints on cosmological parameters by using this method.
With larger collecting areas or longer observing times,
one could probe arcminute scales of the lensing potential and thus
individual clusters. 
We address the effect that foregrounds might have on lensing
reconstruction with 21 cm fluctuations.
\end{abstract}


\maketitle

\section{Introduction}
\label{sec:intro}

The possibility of measuring redshifted 21 cm radiation originating before and during reionization has attracted a lot or interest recently, as it promises new insights into this poorly understood epoch, when the
first radiation sources turned on and eventually ionized the universe. The origin of this observable is the spin flip transition of neutral hydrogen. 
Fluctuations in its brightness temperature can be 
caused by baryonic density fluctuations, inhomogeneities in the spin
temperature, and during reionization also by an inhomogeneous
ionization fraction. Future observations potentially contain
information about all of these 
fluctuation sources \citep{Barkana:2004zy,Loeb:2003ya}. 

These observations could be the definitive probe of  the `dark ages'. At present, there are several probes of the epoch of reionization.  First there is the total Thomson scattering optical depth inferred from the
CMB polarization on large scales points to a mean redshift of reionization, $z_{ri}=10.9^{+2.7}_{-2.3}$ \citep{Spergel:2006hy}. This constraint is prone to large additional systematic uncertainties, as we are just beginning to understand the properties of polarized foreground emission \citep{Page:2006hz}. The CMB offers only an integral constraint so in principle it allows for very complicated evolution scenarios of the ionized fraction, and does not
contain much information as to whether the ionization process is
homogeneous or not.  More detailed polarization information will allow a
few more numbers characterizing the reionization process to be
measured  \citep{Zaldarriaga:1996ke,Hu:2003gh}.

Quasar absorption spectra in the redshift range 5.5-6.5 indicate
a fast evolution of the ionized fraction while $x_H >10^{-3}$ (see e.g. \citealt{Fan:2005es}).
They carry a Gunn-Peterson trough \citep{Gunn:1965hd}, suggesting incomplete ionization of the intergalactic medium (IGM) along the sight line. The sizes of the HII regions directly surrounding the quasars can be used to infer a tighter constraint on
the minimal neutral fraction, $x_H > 0.1$ since the IGM has to be somewhat
neutral otherwise the regions would be larger \citep{Wyithe:2004jw}.

The redshifted 21 cm line is outstanding in comparison to CMB and
quasar observation studies because it should be sensitive to order unity changes in
the neutral fraction $x_H$, hence probing the middle stages of
reionization. It will also provide well-localized measurements along the
line of sight, and it will not require the background of bright
sources which might be rare at high redshifts.

Although 21 cm measurements offer the possibility to
constrain the state
of the IGM during the end phases of reionization, due to the
confounding effects of their unknown astrophysics it might be
difficult to compete with the constraints on cosmological parameters \citep{McQuinn:2005hk}   coming from microwave background experiments such as the Planck satellite
\footnote{http://www.rssd.esa.int/index.php?project=PLANCK\&page=perf\_top},
the E and B experiment (EBEX) \citep{Oxley:2005dg}.
the Atacama Cosmology Telescope (ACT) \footnote{see http://www.hep.upenn.edu/~angelica/act/act.html}
\citep{Kosowsky:2004sw}, or the South Pole Telescope (SPT) \footnote{see
http://astro.uchicago.edu/spt/} \citep{Ruhl:2004kv}).

In this paper we will explore the information contained in the
intrinsically three dimensional 21 cm measurement about the
intervening mass 
distribution at lower redshifts through the lensing effect. We generalize the quadratic estimator technique developed
by \citet{Hu:2001tn} to a three dimensional observable. 
Our formalism uses all the information in both shear and convergence
and properly takes account of correlations along the line of sight. Before the ionized fraction of the IGM becomes substantial, the 21 cm emission against the CMB is a near Gaussian random field. In this regime, which is the focus of the present article, the quadratic estimator should be close to optimal \cite{Hirata:2002jy}.

We will look at 21 cm fluctuations at their lowest possible redshift range,
$z=6-12$, where the second generation of experiments might
be able to measure at high signal-to-noise.
Using our formalism we can also estimate information
losses due to foreground contamination, once these are described by
some model.
We also explore the possibility to constrain
the dark energy density  with lensing
of the 21 cm background. Another application would be to measure nonlinearities in the density field.
We compare our results to the potential of a future high precision
observation of the CMB. The CMB damping tail is an advantage for lensing reconstruction since reconstruction errors decrease with increasing slope, but at the same time it leads to a small scale limitation for CMB reconstruction, as the signal quickly falls below the noise. 

Lensing reconstruction using
the redshifted 21 cm radiation \emph{in absorption} against the CMB
has been investigated by \citet{Cooray:2003ar}, in particular also the
possibility to get a handle on gravity waves from inflation by
gravitational lensing 
cleaning of B mode polarization (lensing converts E to B modes of the
polarization and acts as a contaminant to the primordial signal)
\citep{Sigurdson:2005cp}. Their work describes two types of 
observations, which employ a 20/200 times larger total collecting area than we
will assume here (and five times longer observation time) to
observe angular fluctuations $L_{max}=5000-10^5$ in the 21 cm field. The problem is that the prospect, beating the CMB
level for B mode lensing cleaning, relies on measuring the 21 cm power
spectrum at very high redshifts (they 
use $z_{\rm source}=30$), at which the galactic synchrotron
contamination is a factor $\simeq 20$
larger than for example at redshift $8$. The reason such high redshift
observations 
are needed to compete with likelihood based lensing
estimation is that there is a partial delensing bias when comparing
lenses out to different redshifts and $z_s=30$ turns out to be close
enough to the last scattering surface of the CMB, where gravity waves
are expected to create the B mode fluctuations. These authors furthermore make
the approximation of treating their slices through the 21 cm measurement cube as
uncorrelated, which is not warranted. 

\citet{Pen:2003yv} suggested measuring the effective convergence from
the effect it 
has on the real space variance map of 21 cm fluctuations. The author also
presented rough sensitivity estimates for LOFAR, PAST, and SKA. We improve on this by using all the available information in convergence and shear in an optimal way and use more realistic errors.

In section 2 we will review the theory of the redshifted 21 cm signal in different regimes, which leads to the distinction of a neutral phase where the
brightness temperature fully traces the baryonic fluctuations, as well
as a patchy phase, which can be modeled separately.

In section 3 we review the quadratic estimator technique for lensing
reconstruction following 
\citet{Hu:2001fa}. Then we naturally expand the formalism to the
extraction of weak lensing from an
intrinsically three dimensional signal. The decomposition of the line of sight
component of the signal into modes leads to a hierarchy of independent
lensing backgrounds 
that can be probed with varying precision. Although our concentration lies on applying the quadratic estimator to the epoch immediately before substantial ionization occurs, we scrutinize its applicability to the patchy regime by using an analytic model for the morphology of HII regions.

In section 4 we put our investigation in the context of experiments, estimating
the potential redshift range in which they will observe. Because signal and
noise are evolving with redshift, we break down the volume of
the observation in smaller boxes along the line of sight.
We calculate estimates for the lensing reconstruction future 21 cm
experiments might be able to achieve. 

We present results based on current rough specifications of the Square
Kilometer Array (SKA) in Section 5 and compare them to the possibility of
constraining the matter power spectrum with the CMB temperature and
polarization. The 21 cm approach turns out to have no angular scale
limitation for constraining the convergence.  
We address limitations due to the galactic
foreground and show how these can be incorporated into our formalism
in a straightforward manner. 
We conclude with an outlook and discussion in section 6.

A $\Lambda$CDM cosmology is assumed throughout all calculations, with
parameters 
$\Omega_m=0.3$, $\Omega_\Lambda=0.7$, $\Omega_b=0.04$, $H_0=100 
{\rm h km/sec/Mpc}$ (with h=0.7), and a scale-invariant primordial power
spectrum with $n=1$ normalized to $\sigma_8=0.9$ at the present day.

\section{Theory of the redshifted 21 cm signal}
\label{sec:theory}

The shape of the 21 cm fluctuations is easy to predict in the absence of
radiative sources or 
in a rather homogeneous radiation background. Then
it is simply proportional to the matter power spectrum which
we know based on measurements of the CMB, galaxy surveys, etc. 
In the presence of fluctuations in the ionization fraction, the
observable is more difficult to quantify, due to our lack of knowledge
about the spectrum and efficiency of first sources, and the difficulty of modelling
feedback processes. Some progress has been made using radiative
transfer codes \citep{Sokasian:2003au} on up to 20 $\mpch$ scales, but because of the
large size of the HII regions that simple analytic considerations
suggest (\citealt{Wyithe:2004qd,Furlanetto:2003nf}) it may be helpful for
understanding the large scale morphology of HII bubbles to
also use analytic models, such as the one based on extended
Press-Schechter theory by
\citet{Furlanetto:2004nh}. 

The optical depth of a region of the IGM in the hyperfine transition is
\citep{field59}
\beqa
\tau & = & \frac{ 3 c^3 \hbar A_{10} \, n_{\rm HI}}{16 
k \nu_0^2 \, T_S \, H(z) } 
\label{eq:tauigm} \\
\, & \approx & 8.6 \times 10^{-3} (1+\delta_s) x_H \left[
\frac{T_{\rm CMB}(z)}{T_S} \right] 
\left( \frac{\Omega_b h^2}{0.02} \right) \nonumber \\
& & \times \left[ \left(\frac{0.15}{\Omega_m h^2} \right) \, \left(
\frac{1+z}{10} \right) \right]^{1/2}. 
\nonumber
\eeqa
Here $\nu_0=1420.4 \, \mhz$ is the rest-frame hyperfine transition
frequency, $A_{10} = 2.85 \times 10^{-15} s^{-1}$ is the spontaneous
emission coefficient for the transition, $T_S$ is the 
spin temperature of the IGM (i.e., 
the excitation temperature of the hyperfine transition), $T_{\rm CMB} = 2.73 (1+z) {\rm K}$ is the CMB temperature at
redshift $z$, and $n_{\rm HI}$ is the local neutral hydrogen density.
In the second equality, we have assumed sufficiently high redshifts
such that $H(z) \approx H_0 \Omega_m^{1/2} (1+z)^{3/2}$, which is
well-satisfied in the era we study, $z > 6$.  The local
baryon overdensity is $1+\delta_s=\bar{\rho}/\rho$ and $x_H$ is
the neutral fraction. Here the index in the overdensity emphasizes that
we are measuring quantities in redshift space. The radiative transfer
equation in the Rayleigh-Jeans limit tells us that the brightness temperature of
a patch of the sky (in its rest frame) is $T_b=T_{\rm
CMB}e^{-\tau}+T_S(1-e^{-\tau})$.  Then the
observed brightness temperature increment between this patch, at an
observed frequency $\nu$ corresponding to a redshift $1+z=\nu_0/\nu$,
and the CMB is
\beqa
\delta T(\nu) & \approx & \frac{T_S - T_{\rm CMB}}{1+z} \, \tau 
\label{eq:dtb} \nonumber \\
\, & \approx \, & 26 \, (1+\delta_s) x_H \left( \frac{T_S - T_{\rm
CMB}}{T_S} \right) \left( \frac{\Omega_b h^2}{0.022} \right) \nonumber \\
& & \times \left[ \left(\frac{0.15}{\Omega_m h^2} \right) \, \left(
\frac{1+z}{10} \right) \right]^{1/2} \mK.
\label{eq:tbrightness}
\eeqa
So if the excitation temperature $T_s$ in a region differs from
that of the CMB, the region will appear in emission (if $T_s>T_{\rm CMB}$)
or absorption (if $T_s<T_{\rm CMB}$) against the CMB.
 
\citet{Ciardi:2003hg} argue that the intensity of the Ly$\alpha$
background is large enough at all relevant redshifts, so that
Ly$\alpha$ pumping (the
Wouthuysen-Field effect \citep{wout52,field58},
in combination with the fact that the
color temperature is in thermodynamic equilibrium with the gas kinetic
temperature in an optically thick medium, effectively couples the spin
temperature to the kinetic temperature of the gas. 
Between the CMB last scattering at redshift $z\approx 1000$ and
$z\approx 140$, Thomson scattering of the CMB photons effectively leads
to $T_{\rm CMB}\simeq T_K$, but the gas decouples at $z\simeq
140$ and cools adiabatically until significant structure begins to form.
For the regime we are interested in, $z<15$, it is likely that the
first supernovae and/or
accreting black holes produced enough high energy photons to heat the
IGM \citet{Madau:1996cs}. In this case it can be assumed that a relatively 
homogeneous  
X-ray background leads to $T_S \approx T_K \gg T_{\rm CMB}$
(e.g. \citealt{Ciardi:2003hg,Madau:1996cs}) so that the temperature
factor in Equation \ref{eq:tbrightness} becomes simply unity. In this picture
inhomogeneous effects on the brightness temperature due to shock
heating along sheets and filaments are a minor contribution.

The 21 cm power spectrum is measured in redshift space, hence it is is
related to the actual density contrast power spectrum through 
\beq
P_s(\k)=P_\delta(k) [1+f \mu_k^2]^2
\eeq
where f is the linear growth rate, $f\equiv
\frac{a}{D_1}\frac{dD_1}{da}$. The cosine of the angle between the line of
sight $\hat{z}$ and the wavevector $\k$ is denoted $\mu_k$. 

Here we will assume that in the regime of interest $f\simeq 
1$, hence we obtain with $T_s \gg T_{\rm CMB}$ that the power spectrum of the 21
cm brightness temperature fluctuation in the neutral regime is given
by 
\beq
P_{\Delta T_b}(k) \simeq (26 \mK)^2 \frac{1+z}{10} (1+\mu_k^2)^2 P_\delta(k) 
\label{eq:21pk}
\eeq
In this paper, we shall be mainly concerned with quadratic estimator
lensing reconstruction during this highly neutral regime. The
extension to a patchy epoch is complicated by the presence of a
connected four-point function contribution to the source field. On the
level of the power spectrum this contribution acts as a sample
variance term, correlating different k-modes. In the following
paragraph we shall discuss a semi-analytic model that can be used to
describe the patchy regime. In the analysis section we will briefly estimate
the application of our estimator to this regime.

The fluctuating 21 cm brightness temperature simply traces the matter fluctuations until the time when the IGM starts getting significantly ionized. 
Current radiative transfer codes do not allow
dynamical ranges large enough to simulate the reionization morphology
on scales above $20 \mpch$. To model the patchy regime we implement
the analytic model suggested by
\citet{Furlanetto:2004nh} into the three dimensional signal Gaussian 
random fields of sidelength $1000 \mpch$ comoving (compare
\citealt{Zahn:2005fn}) with a resolution of $512^3$. 
In the
extended Press-Schechter formalism the condition for a region to be
ionized, $\zeta f_{\rm coll.}>1$ leads to a condition on the mean
overdensity within a region of mass m \citep{Furlanetto:2004nh},
\beq
\delta_m \geq \delta_x(m,z) \equiv \delta_c(z)-\sqrt{2} K(\zeta)
      [\sigma^2_\min-\sigma^2(m)]^{1/2}
\eeq
where $K(\zeta)\equiv {\rm erf}^{-1}(1-\zeta^{-1})$, $\zeta$ is the
ionization efficiency factor, $\sigma^2(m)$ is the variance of density
fluctuations on the scale $m$, and $\delta_c(z)=1.686/D(a)$ is the
critical density for collapse scaled to today, with $D(z)$ the scale
factor.
$\sigma^2_\min$ is the variance on the
scale of fluctuations on the scale $m_\min$ the mass for which
$T=10^4 K$ at virialization.

\begin{figure}
\bc
\includegraphics[angle=-90,width=9cm]{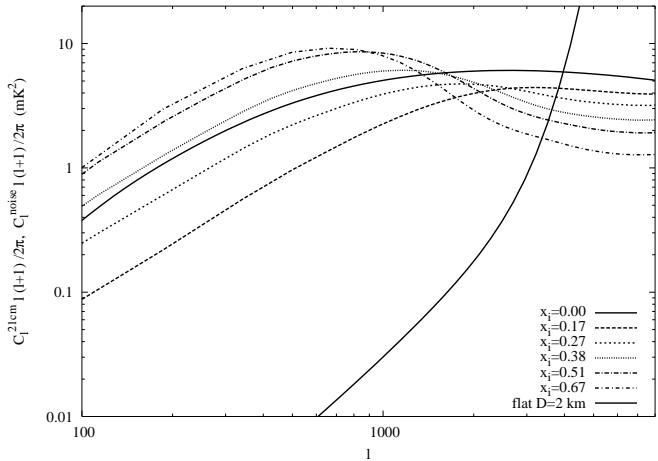}
\ec
\caption{Power spectrum of the 21 cm signal at various stages during reionization. The thick solid line shows the expected noise for SKA for a flat configuration of antennas inside a circle of diameter $2$ km.
When the ionization fraction rises during the expansion of
  HII spheres, first the overall signal decays, then bubbles cluster
  quickly to form large ionized regions tens of $\mpch$ across. This can lead to
  a significant increase in the signal on scales that the first
  imaging 21 cm experiments should be sensitive to. On the smallest
  scales the signal is decreased, making usage of $21$ cm fluctuations for lensing reconstruction somewhat more difficult.}
\label{fig:cl-patchy}
\end{figure}

Figure \ref{fig:cl-patchy} shows the angular power spectrum of 21 cm
fluctuations at
redshift 8 for different values of the efficiency and therefore
ionization fraction. 
We see that up to an ionization
fraction of 
$x_i=0.7$ the signal on small scales decreases (structures below the
bubble scale are washed out) while the ionized regions lead to a bump
in the power spectrum on scales $k \simeq 0.2 \mpch^{-1}$ which
corresponds to an angular multipole at the relevant redshifts of
approximately $l=1000$. At even higher ionized fractions, the entire
patchy power spectrum lies below the neutral case, so it becomes more
difficult to observe the fluctuations and hence to use them for
lensing reconstruction. 
In any event it seems from Figure \ref{fig:cl-patchy} that a
significant part of the patchy regime can be used for 
the lensing reconstruction, in part because the boosted amplitude on
scales $l\simeq 100-1000$ might aid somewhat in this effort. 
We will evaluate these estimates more carefully in Section \ref{sec:compcmb}.
Our choice of ionization efficiency lets reionization begin at redshifts around
8 and evolves rather slowly, due to photon
consumption by clumps in the IGM. It evolves 
through $x_i \simeq 50\%$
at redshift 7 to complete at redshift 6.

We will assume a bandwidth of $B=5\mhz$ for calculating the power
spectrum at various redshifts. This corresponds to a redshift interval $\Delta z=0.286$. During the neutral phase
the density fluctuations evolve slowly, however the sensitivity of the
experiments change rapidly with observation frequency. Because the comoving length scale is given in terms of bandwidth through
\beq
\lbox \approx 1.2 \left( \frac{B}{0.1 \mhz} \right) \left(
\frac{1+z}{10} \right)^{1/2} \left( \frac{\Omega_m h^2}{0.15}
\right)^{-1/2} \mpch,
\label{eq:lcom}
\eeq
our window in frequency space corresponds to a depth of 50 $\mpch$ at
z=6 and to 70 $\mpch$ at z=12. When the first extended HII regions
start forming, the power spectrum evolves more rapidly, however
reionization still only lapses over a comoving length of about 300 Mpc/h, in
comparison to which our window is small. 

We will generate 21 cm brightness temperature power spectra following
Formula \ref{eq:21pk} using linear power spectra that contain the
acoustic oscillation amplitude generated from CMBFAST transfer functions
\citep{Seljak:1996is}. The baryonic wiggles included in the code might
aid somewhat in our 
reconstruction endeavour, for reasons given in the next section. We use the same transfer functions to
implement the signal in three dimensional Gaussian random fields and
model the patchy 
phase.

\section{Weak Lensing Reconstruction}
\label{sec:form}

\subsection{Quadratic Estimator, General Consideration}

Lensing by large scale structure happens whenever there is a
fluctuating background field. At position $\n$ we observe the field
\beq
T(\n)=\tilde{T}(\n+\nabla \phi)
\eeq
where $\tilde{T}$ denotes the unlensed field, $\dtheta=
\nabla \phi$ is the displacement vector while $\phi$ is the
projected potential. Here and in what follows, boldface quantities denote vectors. The projected potential is given in terms
  of the gravitational potential
$\psi(\x,D)$, where $\x$ is position and D is used as a time variable, as
\beq
\phi(\n)=-2 \int d \dang \frac{\dang_A (\dang_s-\dang)}{\dang \dang_s} \psi(D\n,D) \, .
\eeq 
So the power spectrum of the displacements is
$C_L^{\dtheta \dtheta}=L (L+1) C_L^{\phi\phi}$.

An estimator $D(\n)$ for the lensing displacement field information contained in
the temperature field should contain an even number of temperature
terms since the expectation value for odd powers would vanish. It must
also satisfy the condition 
\beq 
\langle D(\n) \rangle =\dtheta(\n)
\label{eq:reconstr}
\eeq
when averaged over many realizations of the background radiation
field, that is for example the CMB or 21 cm radiation.

\citet{Hu:2001fa} showed that the divergence of the
temperature-weighted gradient of the map achieves maximal signal-to-noise among quadratic statistics. 
In Fourier space this quadratic estimator takes the form \citep{Hu:2001kj}
\beq
D_{\rm est.}(\L)=A(L) \int \frac{d^2l}{(2\pi)^2}F(\l,\L-\l)T(\l)T(\L-\l)
\label{eq:dest}
\eeq
The Filter $F(\l,\L-\l)$ is obtained by minimizing the variance of
$D_{\rm est.}$ under the normalization condition for $D(\L)$
\beq
F(\l,\L-\l)=\frac{[C_{l} \L\cdot \l+C_{L-l} \L\cdot (\L-\l)]}{2 \ctot_l \ctot_{L-l}}
\eeq
and the normalization is 
\beq
A(L)=L \left[\int \d2l \frac{[C_{l} L\cdot l+C_{L-l} L\cdot
      (L-l)]}{2 \, \ctot_l \ctot_{L-l}} \right]^{-1}
\eeq
  
Here $\ctot_l$ is the sum of the lensed angular power spectrum of
21 cm fluctuations $\cl_l$ and the noise power spectrum which we will give in
the next section, $C_l^N$. As shown by
\citet{Mandel}, the effect of lensing on the angular 21 cm
power spectrum for an individual plane is small, so one can use the
unlensed power spectrum in place of it. 
The high-pass shape of the quadratic
estimator gives it a property all reconstruction methods share, that they
extract most information from the smallest scales resolved by some experiment. We emphasize that the estimator is unbiased by construction (Equation \ref{eq:reconstr}), independent of Gaussianity of the source field.

With the definition of the lensing reconstruction noise power spectrum
$N_L$
\beq
\langle D^*(\L) D(\L') \rangle = (2\pi)^2 \delta(\L-\L')(C_L^{DD}+N_L)
\eeq
evaluation of the variance of Equation \ref{eq:dest}, $\langle ||D(\L)||^2 \rangle = (2\pi)^2
\delta^D(0) N_L^D$ gives
\begin{eqnarray}
N_L(L) &=& L^2\left[ \int \d2l \frac{[C_{l} \; \L\cdot \l+C_{L-l} \; \L\cdot
      (\L-\l)]^2}{2 \, \ctot_l \, \ctot_{L-l}}\right]^{-1} \\
    &=& A(L) L
\label{eq:hunoise}
\end{eqnarray}
Roughly, when $L^2
C_L^{\phi \phi}=N_L$, then structures down to the angular size $2\pi/L$ in
the lensing field can be reconstructed.

\begin{figure}
\bc
\includegraphics[angle=-90,width=9cm]{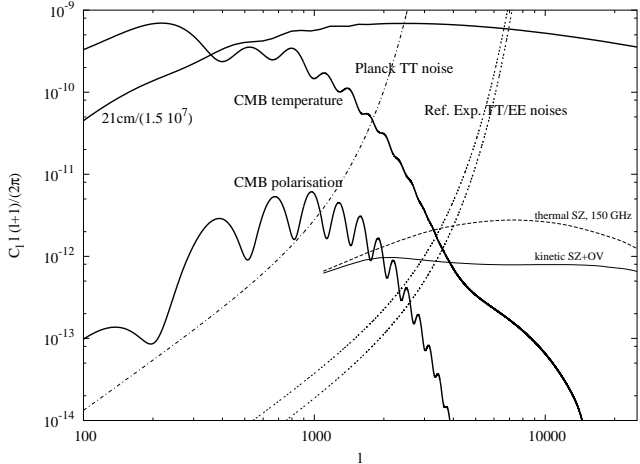}
\ec
\caption{CMB temperature and E polarization power spectrum and noise
  levels of our example experiments: Planck and a futuristic polarized satellite mission (see text for specifications). The 21 cm power spectrum has been rescaled
  to fit into the same plot. The CMB exhibits more pronounced peaks and
  a sharp decay on scales $l\simeq 2500$ above which the
  Sunyaev-Zel'dovich effects become important.}
\label{fig:CMB}
\end{figure}

We illustrate the different power spectrum characteristics of
CMB and 21 cm in Figure \ref{fig:CMB}.

We can investigate qualitatively which properties of the power spectrum 
set the effectiveness of this estimator by looking at the limit $\L
\rightarrow 0$ (on scales where the noise is small and the lensed power is not systematically larger than the unlensed, $\ctot_{l}=\cl_l+\cul_l^N\simeq \cl_l \simeq \cul_l$) 
\begin{eqnarray}
(L^2 N_L)^{-1} &=& \frac{1}{L^4} \int \frac{d^2l}{(2\pi)^2} \frac{[\L \cdot \l C_{l} + \L\cdot
  (\L-\l) C_{L-l}]^2}{2 \ctot_{l} \ctot_{L-l}} \nonumber \\
         &\simeq& \int \frac{dl}{l} \frac{l^2}{2\pi}
  (\frac{1}{2}+\frac{1}{2} \alpha + \frac{3}{16} \alpha^2) + \nonumber \\
   & & +L^2 \int
  \d2l \frac{1}{l^2} g(\alpha,\beta,\gamma) \nonumber\\
   & & 
\label{eq:lowlest}
\end{eqnarray}
where 
\beq
g(\alpha,\beta,\gamma)= \frac{3(5\beta+8\gamma) + \alpha
  (\alpha(2+\alpha)(14+5\alpha+20\gamma))}{384} \, ,
\eeq
and $\alpha\equiv \frac{dln C_l}{dln l}$, $\beta= \frac{d^2
  \ln C_l}{d \ln l^2}$, 
$\gamma = \frac{d^3 \ln C_l}{d\ln l^3}$.
The integrand of the first order term has its
minimum close to a value $\alpha=-2$, in other words when $l^2 C_l$
has no slope. This implies that the CMB with its exponential
decay on small scales will have a lower value of $N_L$ if those scales
are resolved. Integrating Equation \ref{eq:lowlest} to the maximum
multipole $l_\max$ at which $C_l^S > C_l^N$ leads to
\begin{eqnarray}
(L^2 N_L)^{-1} &\simeq& \frac{1}{2\pi}(\frac{1}{2} +\frac{1}{2} \alpha
  + \frac{3}{16} \alpha^2) \frac{l_\max^2}{2} \\
      & & + \frac{L^2}{2\pi}
  g(\alpha,\beta,\gamma) \ln(l_\max/l_\min) \, , \nonumber\\
& &
\label{eq:intlowlest}
\end{eqnarray}
where $l_\min$ is the lower bound of the integration. The second term
will lead to departures from a constant value of $(L^2 N_L)$. The
scale at which both terms are comparable is 
\beq
L_{\rm comp.} \simeq \frac{l_\max}{\ln
  (l_\max/l_\min)^{1/2}}\sqrt{
\frac{1+\frac{1}{2}\alpha +
  \frac{3}{16}\alpha^2}{g(\alpha,\beta,\gamma)}} \, .
\eeq 
Evaluation of this expression shows that for a sloped power spectrum $l^2C_l$,
such as that of the CMB, $L_{\rm comp.}$ is lower than for a spectrum
that is nearly constant, such as that of 21 cm fluctuations.
The second order term contributes strongly when the slope is
large. If the measured power spectrum has a small 
slope, as is the case with 21 cm fluctuations,  $L^2 N_L$ can be
expected to be nearly constant to the scale $l_\max$ where noise
becomes important.  

In the following section we will generalize the quadratic estimator to a three
dimensional observable that is used to reconstruct the lensing
field. The final estimator is then simply the sum of 
estimators of k modes along the line of sight. 
This allows one to improve the constraints above those by the CMB especially
on small scales by summing over sufficiently many lensing backgrounds.

For our analysis in Section \ref{sec:exp} we will compute the
angular power spectrum of the deflection angle as the integral over line of sight 
\begin{eqnarray}
C^{\dtheta \dtheta}_L = \frac{9 H_0^4 \Omega_0^2}{L (L+1) c^2} \int_0^{D_s}   & & dD_l
\left(\frac{D_{ls}}{D_s \, a(D_l)}\right)^2 \times  \nonumber \\ & & \times P_\delta(k=\frac{L}{D_l}, D_l)
\end{eqnarray}
over the power spectrum of mass fluctuations. $D_x$ denote angular
diameter distances with $x=l,s$ for lens and source. 

We use the halo model fitting function to numerical simulations of
\citet{Smith:2002dz} to generate nonlinear $\Lambda$ CDM power
spectra as input for the deflection angle integral.

\subsection{Extension to a three dimensional signal}

Different from the CMB, in the case of 21 cm brightness
fluctuations we will be able to use multiple redshift information to
constrain the intervening matter power spectrum. 
One could imagine applying this estimator to succesive planes
prependicular to the line of sight. However the different planes would
be correlated and there is no straightforward criterion to take this
correlation into account in establishing the final estimator, since it would depend on redshift, source properties, and during reionization on the mean
bubble size and distribution.
Instead we will use the knowledge of the three-dimensional information
in a different way, dividing the temperature fluctuations in Fourier
space in fluctuations $\k_\bot$ perpendicular to the line of sight,
and a component $k_\|$ in the frequency direction.

We devide a volume on the sky to be probed by a given 21 cm survey
into a solid angle $d \Omega$ and a radial coordinate $z$. Components
of wavevectors along the line of sight are described by $k_{\|}$ and
those perpendicular to the line of sight are given by the vector
$\k_\bot$, which is related to the multipole numbers of the spherical
harmonic decomposition on the sky as $\l=\k_\bot \dang$, where $\dang$ is the
angular diameter distance to the volume element we are probing.

Suppose we want to measure a field $I(\r)$ with power spectrum $P(\k)=P(\k_\bot,k_\|)$. Converting $k_\bot$ to
angular multipole

\begin{eqnarray}
I(\r)&=&\int\frac{d^3k}{(2\pi)^3} I(\k) e^{i\k\cdot \r} \\
     &=&\int \frac{d^2l}{(2\pi)^2} \int \frac{dk_\|}{2\pi}
     \frac{I(\k_\bot,k_\|)}{\dang^2} e^{i(\l \cdot \theta + k_\|z)} \nonumber
\end{eqnarray}
with $\tilde{I}(\l,k_\|)=\frac{I(\k_\bot,k_\|)}{D^2}$ we have
\begin{eqnarray}
\langle \tilde{I}(\l,k_\|)\tilde{I}(\l',k_\|') \rangle&=&\delta^D(\k-\k') (2\pi)^3 \frac{P(\k_\bot,k_\|)}{\dang^4} \nonumber \\
&=&(2\pi)^2\delta^D(\l-\l')(2\pi)\delta^D(k_\|-k_\|') \times \nonumber \\
& & \times \frac{P(\k_\bot,k_\|)}{\dang^2}
\end{eqnarray}
where in the second step a factor of $\dang^2$ got absorbed into $\delta^D(\l-\l')$ because of $\l=\k_\bot \dang$.

Let us discretize the z direction of a real space observed volume with
radial length $\lbox$,
\beq
k_\|=j\frac{2\pi}{\lbox} , \quad \delta^D(k_\|-k_\|')=\delta_{j_1j_2}(\frac{\lbox}{2\pi})
\eeq
so that
\beq
I(\r)=\int\frac{d^2l}{(2\pi)^2} \sum_j \left(\frac{I(\k_\bot,k_\|)}{\dang^2
  \lbox}\right) e^{i\k\cdot\r}
\eeq
It makes sense to define
\beq
\hat{I}\equiv \frac{I (\k_\bot, k_\|)}{\dang^2 \lbox}
\eeq
so that on the sphere we have
\beq
\langle \hat{I}_{j_1}(\l_1)\hat{I}^*_{j_2}(\l_2)\rangle = (2\pi^2)\delta^D(\l_1-\l_2) \delta_{j_1j_2}\left(\frac{P(k,\mu_k)}{\dang^2\lbox}\right)
\eeq
where $\mu_k$ is the cosine between the wave vector and the line of sight. 

We have for the angular power spectrum for separate values of j that (including redshift space distortions)
\beq
C_{l,j} \equiv (1+\mu_k^2)^2\frac{P(\sqrt{(l/\dang)^2+(j\, 2\pi/\lbox)^2})
}{\dang^2 \lbox} \, ,
\label{eq:pk-to-cl}
\eeq
where P now represents the spherically averaged power spectrum. We have also introduced the notation 
$C_{l,j}$, denoting the power in a mode with angular component l and radial component $k_j=j \, 2 \pi/\lbox$.

We show in the Appendix that because modes with
different j can be considered independent, the best estimator can be
obtained by combining the individual estimators for 
separate j's without mixing them (in the sense of making quadratic
combinations of them). As long as the IGM is not substantially ionized, the assumption of Gaussianity is justified at the redshifts of interest, $z\simeq 6-12$, where non
linearites in the gravitational clustering are small on observed scales of
several Mpc/h.
We find the three dimensional lensing reconstruction noise defined by
\beq
\langle D(\L) D^*(\L')\rangle = (2\pi)^2 \delta(\L-\L') N_L^D
\eeq
to be (Equation \ref{eq:noisesum} of the Appendix, where e.g. $\cul_l \rightarrow  \pul_l$)
\begin{eqnarray}
N_L^{D} &=& \frac{1}{\sum_k \frac{1}{L^2}\int \d2l \frac{[\cul_{l,k}
    \L\cdot \l + 
    \cul_{L-l,k} \L\cdot (\L-\l)]^2}{2 \ctot_{l,k}\ctot_{L-l,k}}} \\ 
      &=& \frac{1}{\sum_k N_{L,k}^{-1}}
\end{eqnarray}
where in the last line we have just substituted the standard
expression, Equation \ref{eq:hunoise}. Note that analogous to the CMB case, $\ctot_l=\cl_l+C_l^N$, where $C_l^N$ is the noise power spectrum, and $\cl_l \simeq \cul_l$ in the case of 21 cm fluctuations.

If there is a connected four point function contribution during the
epoch of extended HII regions, 
this will add a term to the variance of the estimator. This will
change the weightings $A_L$ as well, and make the estimator suboptimal. We will
not try to come up with an estimator that is fully compatible with the
non Gaussian signal due to patchy reionization. Instead, in the next
section we will conservatively estimate that sensitivities get worse by a factor
$\frac{1}{\sqrt{x_H}}$, where $x_H$ is the ionized fraction. In other words, we will
  effectively be treating the patchy regions as part of the source field
  which can be masked.

\begin{figure}
\bc
\includegraphics[angle=-90,width=9cm]{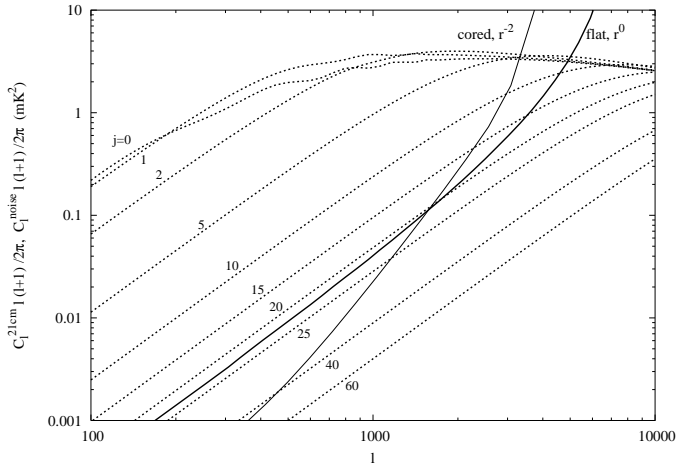}
\ec
\caption{The dotted lines show the angular power spectrum for different $k_{\|,j}=j \, 2 \,
  \pi /L$, labeled by $j$. Going to higher values of $j$, the signal decays quickly below S/N=1 and only the first 15 or so
  modes contribute to the final lensing estimator. This number depends of course on the thickness of the redshift interval probed (in this case $B=5 \mhz$, corresponding to $\Delta z=0.286$ at $z=8$). The thick and thin solid curves are for a flat and a cored antenna configuration of SKA (see the text).} 
\label{fig:clnoise}
\end{figure}

Let us examine the contribution of the components $C_{l,j}$ to the total
noise in the estimation of the deflection field. From Equation \ref{eq:pk-to-cl} we see that with higher values of the
component $k\|$, higher values of the three dimensional power spectrum
$P(k)$ will translate into each $C_l$ value. 
But $P(k)$ is monotonically falling on all scales of interest
(foregrounds can be expected to contaminate constraints
below ${\rm k}<10^{-2} {\rm h/Mpc}$, the scale of the horizon at matter radiation
equality and the turnover of the power spectrum), so that
effectively the angular power spectrum amplitude will drop with $j$
towards the noise level.
We show this in Figure \ref{fig:clnoise} for our bandwidth choice of $5 \mhz$.
Notice that on the smallest resolved scales, $l \simeq 2000-5000$ the
signal increases slightly as we go from the fundamental $k_\|=0$ to
the next higher modes in our line-of-sight decomposition. This is
because the radial component of the power spectrum is increased due to
redshift space distortions. Because the smallest resolved angular scales
contribute most to the lensing reconstruction, in case of the first
few modes in the $k_\|$ decomposition this increase overcompensates for
the general decay of $C_{l, j}$ on the smallest scales.
However overall for each rectangular data
field of a given bandwidth we will only be able to sample a limited number of modes $k_\|$ with
signal-to-noise greater than one, this number being proportional to the
frequency depth/bandwidth of the field.

The 21 cm power spectrum and, as we will see in the next section, the
noise of the experiment depend on redshift, so we calculate them for
volume elements corresponding to the above choice a $5 \mhz$ bandwidth. Beginning at
the end of reionization, each volume element 
contributes to reducing the reconstruction noise while we go along,
until the signal-to-noise of the experiment
at high redshift becomes negligible.  We should wonder whether long wavelength
modes overlapping neighbouring volumes lead to an underestimation of
the final lensing noise, 
as we are treating each volume element as uncorrelated.
We made a simple test by comparing the $N_L$ of a pair of neighbouring
volumes to the sum of 
$N_L$'s of each element. With our choise of a $5\mhz$ bandwidth we
found no excess information $>1\%$ in the sum of the seperated field's
$N_L$'s, meaning that we can safely neglect those correlations.
A multiplicity of those volumes can thus be used to reconstruct each
lens just by summing over them.

\section{Antenna Configuration, Sensitivity Calculation}
\label{sec:exp}

We have found out in the previous section that in order to measure the lensing
signal with redshifted 21 cm fluctuations, we need small angular scale
resolution as well as wide redshift coverage. Sensitivity calculations for 21 cm experiments are presented in \citet{Morales:2004ca} and \citet{McQuinn:2005hk}. The sensitivity diminishes quickly with rising foreground temperature at longer wavelengths. The foreground temperature is dominated by galactic synchrotron.
To observe at high redshift one needs to compensate for the increased foreground by
increasing the collecting area or number of antennas. The angular resolution is improved
by increasing the maximal baseline. To have good angular Fourier mode
coverage we want all baselines to be represented though. Hence an
increase in angular resolution has to be compensated for by increasing the collecting area
in order to keep the covering fraction the same. 
In this section we will discuss various aspects of the first two
generations of 21 cm experiments in the context of using them for
lensing reconstruction.

A standard approach in
radio astronomical measurements (see e.g. the Very Large Array (VLA) \footnote{http://www.vla.nrao.edu/},
the Low Frequency Array LOFAR \footnote{http://www.lofar.org/}, and
the Atacama large millimeter array (ALMA) \footnote{http://www.alma.nrao.edu/} configurations) is to
have a power law decay in the number density of antenna $\mathcal{P}\propto r^{_\alpha}$ with 
radius. This translates into a drop in the
number of baselines with separation. The distribution flattens out
towards the center to $n(r)=r^0$ simply because antennas cannot be
stacked closer to each 
other than their individual physical size. We found in the previous
section that the contribution to the weak lensing estimator of $C_{l,j}$
becomes smaller quickly with higher line of sight modes $j$ used. This is
just an expression of the fact that what we are interested in is
displacements of 21 cm photons perpendicular the line of sight. Lensing
reconstruction works through a large number of individual sources
being aligned around a big deflector, hence the imaging of small
scales is important. For probing small angular scales a flat array
profile (that is without a power law 
decay) may offer an advantage, especially if
the observable angular power spectrum falls slightly beyond
$l\simeq 1000$ and we want to probe it on those scales. 

We can calculate the number of baselines as a function of visibility u
from performing the autocorrelation
\beq
n(u) = \lambda^2 \int d^2r \, \mathcal{P}_{\rm ground} (\r+\x) \,
\mathcal{P}_{\rm ground} (\r)
\eeq
where $\lambda$ is the observed wavelength and $\mathcal{P}(r)$
denotes the radial profile of the circularly symetric 
antenna distribution. $\x$ is the vector of seperation of an antenna
pair, $\x=\lambda \u$.

The time a particular visibility $u$ is observed, $t_u$ is given by
\beq
t_u= \frac{A_e t_0}{\lambda^2} n(u)
\label{eq:t-of-u}
\eeq
where $A_e$ is the
effective antenna area and $t_0$ is the
total observing time. We will assume 2000 hours for our calculations, which 
might be achieveable with planned observatories within a single seasons.

The sensitivity for a given array distribution and specifications can
then be calculated as follows. The RMS fluctuation of the thermal noise per
pixel of an antenna pair is \citep{White:1997wq,Zaldarriaga:2003du}
\beq
\Delta T^N(\nu) = \frac{\lambda^2 B T_{\rm sys}}{A_e\sqrt{B t}}
\eeq
where $B$ is the bandwidth of the observation. 
For a single baseline the thermal noise covariance matrix becomes
\beq
C_{ij}^N=\left(\frac{\lambda^2 B T_{\rm sys}}{A_e}\right)^2\frac{\delta_{ij}}{Bt_\u}
\eeq
where $T_{\rm sys}$ is the system
temperature, dominated by galactic synchrotron radiation (roughly,
$T_{\rm sys}\propto \nu^{-2.55}$ with $T_{\rm sys}=440K$ at $z=6$) and $B$ is the
bandwidth of this frequency bin of the total observation.
From this we get the noise versus angular multipole number because of
$2 \pi u=l$ through $\frac{d^2l}{2\pi} C_l^N = d^2u C_u^N$.

We are now going to asses the potential of planned 21 cm experiments
to measure the lensing imprint using this quadratic estimator. There
are currently four major 
experiments under way, the Mileura Wide Field array 
(MWA) \citep{Morales:2004ca}, the primeaval structure telescope \citep{Pen:2004de}, the Low Frequency Array (LOFAR) \footnote{http://www.lofar.org} and in
the second generation planning stage, the Square Kilometer Array (SKA)
\footnote{http://www.skatelescope.org/}. While these projects differ
qualitatively in the hardware used, the most crucial difference is their total
collecting area. For SKA, current proposals call for 5000 antennas with an individual collecting area
of $120 m^2$ at $z=8$ (the effective antenna area depends on
wavelength and hence redshift through the square).  The collecting
areas for MWA and LOFAR/PAST are significantly smaller, 1\% and 10\%
of that of SKA respectively.  The optimal array can be
designed by distributing the total collecting area over a large number
of antennas. The number of baselines goes as $N_{ant}^2$ which enters
into Equation \ref{eq:t-of-u}. In other words, since an array with say
ten times the number
of dishes (each ten times smaller) as SKA has a ten times larger survey speed, it is better suited
for EOR observations. However the computational requirements for the
associated correlator unfortunately also scale as $N_{ant}^2$.

We show the result of our imaging sensitivity calculation for a flat
($r^0$) to $r_{\rm max}=1500$ m and a cored ($r^0$ to $r=80$ m, then
$r^{-2}$ to $r_{\rm max}=1500$ m)
array of SKA antenna in Figure \ref{fig:clnoise}, together with the
hierachy of decreasing angular power 
spectra as the line of sight component of the 3D measurement $k_\|$ is
decreased. We find that a constant radial density
of antennas turns out to offer the best compromise between the number of $(\k_\bot,k\|)$ modes
that can be probed and overall angular resolution. 
For a cored
$r^{-2}$ distribution for example, the improved sensitivity on those
intermediate scales does not compensate for the loss in angular
resolution that comes from the lack of large baselines relative to a
flat distribution. 

For statistical detections of the
convergence on the sky a large field of view is desired. The angular
power spectrum errors go as $1/\sqrt{f_{\rm sky}}$. Hence it is
advantageous to distribute the same total collecting area in many
smaller antennas, thus increasing the speed of the survey. 
On the down-side, it is more difficult to scale small dipole like 
antenna to high freqencies (below the redshift of reionization), where
other interesting science with 21 cm emitters lies
(see e.g. \citealt{Zhang:2005pu}).

\section{Results and a Comparison with the CMB}
\label{sec:compcmb}

In this section we will establish a specific configuration that seems
likely within reach of the second generations of EOR observatories and
calculate our three dimensional lensing estimator for a particular
reionization scenario. This way we will gauge the prospects of lensing
reconstruction using redshifted 21 cm fluctuations and how they relate
to using other backgrounds.

The prospects of lensing reconstruction with the quadratic estimator have been
explored in depth by
\citet{Hu:2001fa,Hu:2001kj,Hu:2001fb,Amblard:2004ih} in the
context of the CMB.  A systematic problem
with temperature is that it is contaminated by Doppler related
anisotropies (the kinetic Sunyaev-Zel'dovich effect 
\citealt{Sunyaev:1980nv}) which share the same 
frequency dependence as the primordial CMB. \citet{Amblard:2004ih}
showed that this contamination is significant, and might eventually
have to be taken care of by masking thermal SZ detected areas from the
map. It is found that using the CMB 
polarization an order of magnitude improvement over
temperature reconstruction might be possible \citep{Hu:2001kj}.
On scales where the B component of polarization can be resolved (above
$l\simeq 100$ this component is almost entirely produced by gravitational
lensing rotation of the E modes), with iterative likelihood techniques
astonishingly there may be no limit at all to delensing
\citep{Seljak:2003pn}, on angular scales where the B modes can be resolved.  

As discussed in the previous section (Equation \ref{eq:lowlest}), the
effectiveness of lensing 
reconstruction depends on the slope of the power spectrum, and
the 21 cm background suffers a disadvantage compared to the 
CMB in that there is no damping tail and the traces of baryonic
oscillations are comparatively small. On the other hand the lack of an
exponential decay in the 21 cm fluctuations suggests that these can be
used for reconstruction out to smaller scales.
Figure \ref{fig:CMB} shows CMB strengths and limitations
in comparison to the shape of the 21 cm power spectrum (which has been
rescaled to fit in the same plot). 

The damping tail of the CMB at scales $l \geq 3000$ leads to a limit for the
deflection scales $l\simeq 1200$ that can be probed using quadratic
techniques \citep{Hu:2001kj}. 
The 21 cm power spectrum does not decrease substantially at small
scales, the only theoretical limit being set by the Jeans scale.    

\begin{figure}
\bc
\includegraphics[angle=-90,width=11.7cm]{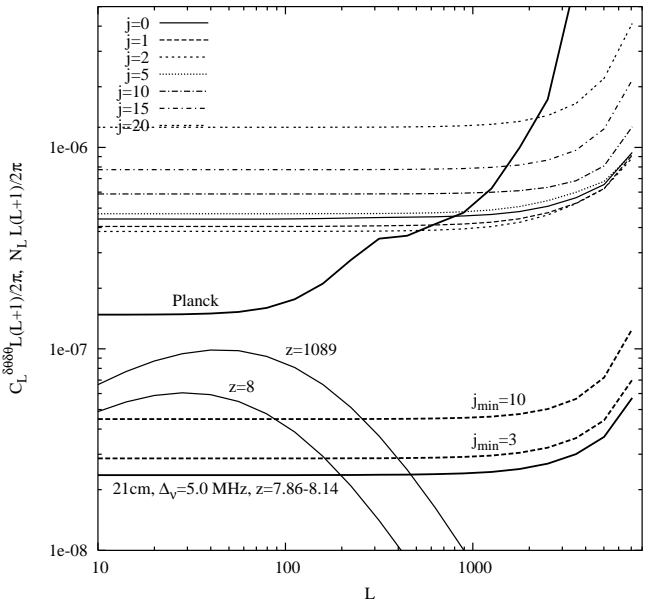}
\ec
\caption{Lensing reconstruction noise $N_L$ for one
  redshift interval centered at $z=8$ corresponding to a bandwidth of
  $5\mhz$. The curves labeled `z=8' and `z=1089' are the displacement field power spectra for 21 cm and CMB as source respectively. The thick solid curves labeled `Planck' and `21 cm' are the lensing reconstruction noises we find. The 21 cm noise is based on combining all $k_\|$ modes. The thin lines labeled `j=0-20' on the other side are the results for individual $k_\|$-modes. We see that from this redshift range alone the combined temperature and
  polarization information of Planck can be beaten. The necessity to
  substract foregrounds lessens the constraint from 21 cm somewhat. They effectively render the first few $k_\|$ modes useless for the reconstruction, see the text. The resulting noise levels are shown in the thick dashed curves labeled  $j_{\rm min}=3$ and $j_{\rm min}=10$, for a less and more conservative assumption about the complexity of foreground contamination respectively.}
\label{fig:NL-hierarchy}
\end{figure}

Figure \ref{fig:NL-hierarchy} shows a rather constant shape of $L^2 N_L$ versus
the angular multipole of the lensing field $L$. 
This is essentially a consequence of the rather flat shape of the 21 cm
power spectrum, as we discussed following Equation
\ref{eq:lowlest}: in this case we would expect $N_L$ to not change
much up to scales where the signal becomes comparable to the noise of
the experiment.
This property will allow us to image the lensing field
down to small scales. 
We also show in the same figure a plot
of the temperature based CMB lensing reconstruction based on the
Planck satellite experimental specifications.

The CMB power spectrum sensitivity is given in terms of detector noise
$w^{-1}$ and
beam $\sigma_{\rm FWHM}$ \citep{Knox:1996cd}
\beq
C_l^{\rm noise} = w^{-1} e^{l (l+1) \sigma_{\rm FWHM}^2/8 ln 2}
\eeq
We assume two types of experiments, Planck, with $w^{-1/2}=27 \mu \K-\arcmin$ at $5 \, \arcmin$ angular resolution, and a futuristic experiment
with noise level $w^{-1/2}=3 \mu \K-\arcmin$ at an angular resolution of $3\, \arcmin$.
For the polarization we use as usual that $w^{-1/2}_P=\sqrt{2} w^{-1/2}_T$
if all detectors are polarized. We include the noise power spectra for
Planck temperature, and for 
polarization and temperature measurements of our reference CMB
experiment in Figure \ref{fig:CMB}. 

We calculate the minimum variance lensing noise level following
\cite{Hu:2001kj} 
\beq
N_{\rm mv}(L) = \frac{1}{\sum_{\alpha,\beta}({\rm N}^{-1}(L))_{\alpha \beta}}
\eeq
where $\alpha$ and $\beta$ run over T, E, and B temperature and
polarization fluctuations. The noise levels $N(\L)_{\alpha
  \beta}$ take on different forms depending on whether $\alpha$ and
$\beta$ are equal (the
$BB$ term is generally considered to have vanishingly small signal to
noise), or different, $\alpha \beta = \theta E$, $\theta B$, and
$E B$.  

\begin{figure}
\bc
\includegraphics[angle=-90,width=11.7cm]{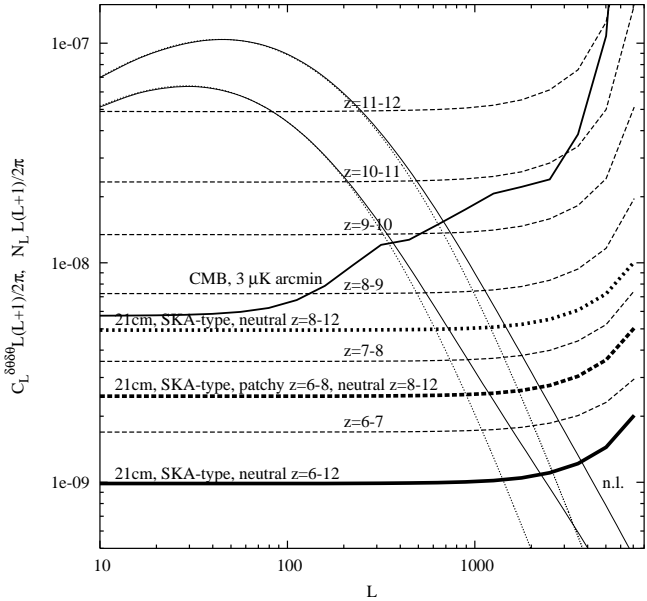}
\ec
\caption{Lensing reconstruction noise with an experiment that
  has the total collecting area of SKA, if the IGM is not ionized
  during the regime z=6-12, shown in the thick line at bottom. The
  individual contributions from redshift intervalls are shown in the dashed
  lines. The noise levels are compared to the deflection angle
  power spectrum, where the solid (dotted) lines are for the nonlinear
  (linear) density fluctuations at z=8 and z=1089. On scales of $L=1000$ and
  above, our method might be able to achieve an order of magnitude
  lower total noise levels than what is 
  possible with the CMB quadratic estimator technique (shown in the
  other thick solid line). Here the
  smallest angular scale reconstructed using EOR fluctuations is
  $L_{\rm max}=2250$. The thick dashed line shows the increased noise
  level if reionization is inhomogeneous during $z=6-8$, calculated in the way
  detailed in the text. Finally the thick dotted curve gives the total noise
  level if the regime $z=6-8$ is not used at all in the analysis.}
\label{fig:NL-z-hierarchy}
\end{figure}

We compare Planck constraints to 21 cm observation sensitivities for
the deflection 
angle power spectrum that we get from observing a $5\mhz$ slice at redshift 8
in Figure \ref{fig:NL-hierarchy}. The corresponding redshift depth is
$\Delta z =0.5$. If we want to gather all the information accessible
to us from fluctuations in 
the 21 cm brightness temperature, we can add up a number of redshift intervals
with the constant bandwidth, employing the whole redshift range
covered by the observation. We find that even when using only a fraction of the entire data volume available in 21 cm, $\Delta z \simeq 0.3$, Planck's combined temperature and polarization potential for doing lensing reconstruction can be beaten with the three dimensional generalization of the quadratic estimator.
If we want to use a larger redshift range for the 21 cm reconstruction, we need to take into account the redshift evolution in the power
spectrum of matter fluctuations, and that of the ionization
fraction. For an SKA 
type experiment in the optimized configuration we use, this range is
$z=6-12$, at 
higher redshifts the sensitivity becomes too small for imaging, the limitation being set by the increased temperature of the foreground. Note that if reionization completes earlier than $z=12$ lensing reconstruction using 21 cm will be impossible with SKA. In our particular reionization scenario
we achieve a constraint 
at $L\approx 1000$ about ten times as good (i.e. an $N_L$ ten times lower)
for the lensing 
reconstruction, in comparison to our reference CMB experiment with 3
arcminute resolution 
and a noise level of $w^{-1/2}=3\mu \K-\arcmin $. We show this result in
Figure \ref{fig:NL-z-hierarchy}, together with nonlinear/linear
lensing field power spectra. The Figure also shows (in the thick dashed 
curve) the increase in noise when our patchy model is used for
the range 6-8 in place of an extension of the neutral phase. 
This increased noise relative to the case of probing the
  pre-reionization IGM has two reasons: first, as shown in Figure
  \ref{fig:cl-patchy}, the fluctuation level of the 21cm signal is
  decreased on the smallest resolved scales, making this source less
  valuable for lensing reconstruction. Secondly, the connected
  four-point function contribution adds a non Gaussian term to the
  noise covariance matrix of the power spectrum, acting as a sample
  variance term in correlating different band-powers. We treated this
  term in a simplified manner by assuming that the majority of bubble
  features arises on scales well above the resolution scale of SKA,
  about 1'. Then the bubbles can be resolved and masked when
  establishing the final estimator. The mask increases the sample
  variance simply as $\delta C_l^*=\frac{1}{\sqrt{x_H}}\delta C_l$,
  where $x_H$ is the neutral fraction at this redshift. Our estimate
  of the lensing reconstruction noise for the patchy epoch is
  conservative, in that power 
  in the regions outside the bubble mask will not be suppressed, however we
  use the global average power spectrum which is suppressed by
  $x_H^2$. For completeness we also show (in the thick dotted curve)
  the total lensing reconstruction noise if we only use the neutral
  regime above $z=8$ in the analysis.

\begin{figure}
\bc
\includegraphics[angle=-90,width=9cm]{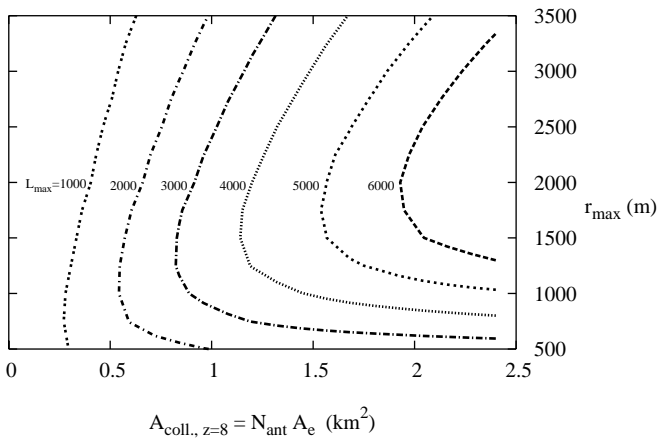}
\ec
\caption{Contour plot of $L_{\rm max}$, the largest displacement field multipole probed. From this Figure one can infer the optimal array radius (assuming here
  an $r^0$ distribution of antennas) that one should choose
  given a total collecting area $A_{\rm coll}$. It also shows that if one
  would have four times the collecting area of SKA (for which
  $A_{\rm coll., z=8}=0.6 {\rm km}^2$), or if one were to observe
  four times longer on the same patch of the sky, very high multipoles
  could be probed.} 
\label{fig:nant-rmax-lmax}
\end{figure}

Because of the multiple background information, in reconstructing the
deflection field we can
approach the limit imposed by the finite angular resolution of an experiment
(determined by its maximum baseline). 
In Figure \ref{fig:nant-rmax-lmax} we show contours for the maximum 
lensing deflection field multipole $L$ that can be probed as a function of
the maximum baseline and the total collecting area. For the $r^0$ 
array there exists an optimal $r_{\rm max}$ which depends on the amplitude of the power
spectrum and the total collecting area. If one uses a wider radius
while keeping the number of dishes constant, the noise level increases
on all angular scales, so although one might be able to use slightly
larger multipoles (smaller scales) for the reconstruction, many modes
of our hierarchy do not enter the estimator. On the other hand, if one
decreases the array size, thus making the antenna distribution denser,
this leads to a lower noise level on relatively large scales, while no
baselines exist anymore to measure the small scales that are crucial
for the reconstruction. 

We find that after gathering 2000 hours of data, a very ambitious
experiment with four times the 
collecting area of SKA (for which $A_{\rm coll., z=8}=0.6 {\rm km}^2$)
could detect lensing at scales beyond $L=6000$, which is comparable to
the scale of galaxy clusters.

We will extend considerations to statistical detections of the
deflection power spectrum. 
Because the geometrical ratio $\frac{D_l D_{ls}}{Ds}$ is larger for
the CMB (qualitatively the bulk of 
lensing happens at angular diameter distances that are closer to the
middle between the observer and the CMB), the respective deflection angle power
spectrum has a higher amplitude. On the other hand the 21 cm
experiment will have the advantage of measuring multiple planes so that at
comparable angular resolution smaller errors in the angular power spectrum at high $L$ can in principle be achieved. 

In \citet{Sigurdson:2005cp} it was proposed that 21 cm
reconstruction of the lensing power spectrum might be helpful in
getting at B mode polarization from primordial gravity waves. Since lensing
partially converts E (gradient) polarization into a curl
component, this secondary signal swamps the B modes from a possible
inflationary tensorial fluctuation background. The problem is that at
significantly lower redshifts than the last scattering surface, only
part of the lensing structure encountered by the CMB photons is
traced, leading to a delensing bias if 21 cm fluctuations are
used. One would have to observe at high enough
redshifts if one were to compete with CMB polarization
experiments. Indeed the authors find that in principle with an ultra
sensitive experiment (for instance space based) observing at redshift
30 one might be able to beat CMB limits beyond the iterative
likelihood approach of \citet{Seljak:2003pn,Hirata:2003ka}.
To achieve this one would need 1-2 orders of magnitude more
collecting area than what is planned for the second generation of
observatories, hence this application is beyond the scope of our paper.

\begin{figure}
\includegraphics[angle=-90,width=9cm]{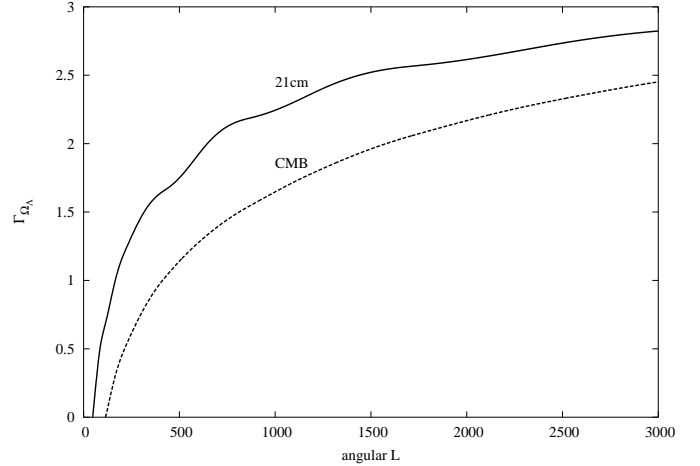}
\caption{Plot of the parameter $\Gamma$ defined in the text. This
  suggests that 21 cm fluctuations are better suited in principle
  to measure 
  the value of the cosmological constant than the CMB, assuming that the same
  number of angular 
  modes can be probed.}
\label{fig:ratio-OL}
\end{figure}

A characteristic of 21 cm fluctuations is that the distance between
observer and lens is a larger fraction of the whole distance to the
source, so one would imagine that if the same number of multipoles
are probed with 21 cm as with the CMB (by having a lower noise level),
one would obtain different constraints on dark energy. We show this in
Figure \ref{fig:ratio-OL}, where the ratio  
\beq
\Gamma_{\Omega_\Lambda} (L) = \frac{\Omega_\Lambda}{C^{DD}_L} \frac{\partial
  C^{DD}_L}{\partial \Omega_\Lambda}
\eeq
is plotted against $L$. The quantity measured
shows how well the dark energy density can be constrained by using
information 
from a given angular scale. Its value gives the constraint that can be
put on the parameter from this scale if there were no degeneracies with
other unknowns. We see that when the same range of angular scales
are resolved, the 21 cm fluctuations fair somewhat better in constraining the dark
energy density $\Omega_\Lambda$.

The noise for estimation of bandpowers is reduced by averaging over
$\L$ directions in a band of width $\Delta L$
\beq
\Delta C_L^{DD} =\sqrt{\frac{2}{(2L+1)\Delta L f_{\rm sky}}} [C_L^{DD} +
  N_{\rm mv}(L)] \, .
\eeq

\begin{figure*}
\bc
\includegraphics[angle=-90,width=8.5cm]{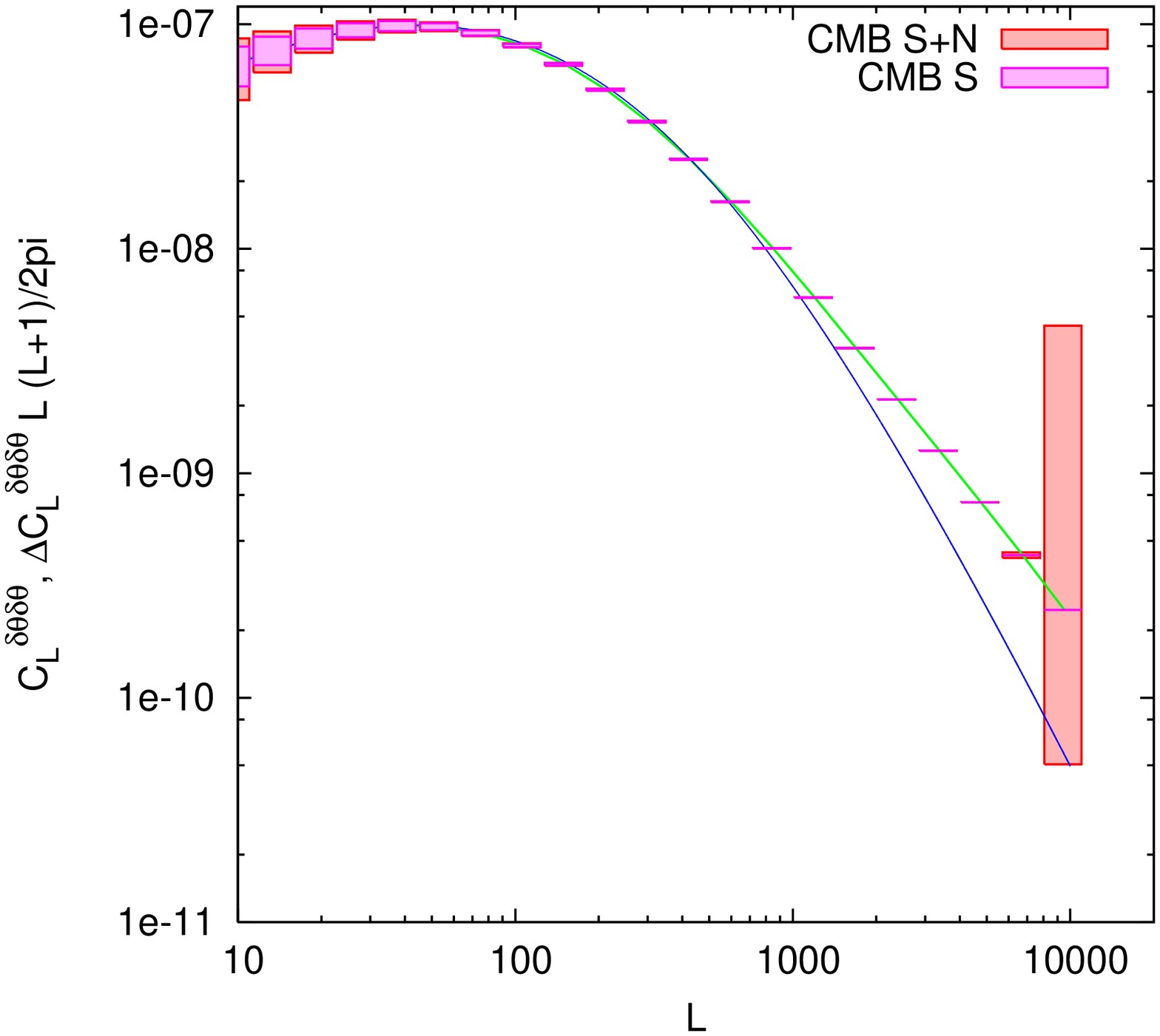}
\includegraphics[angle=-90,width=8.5cm]{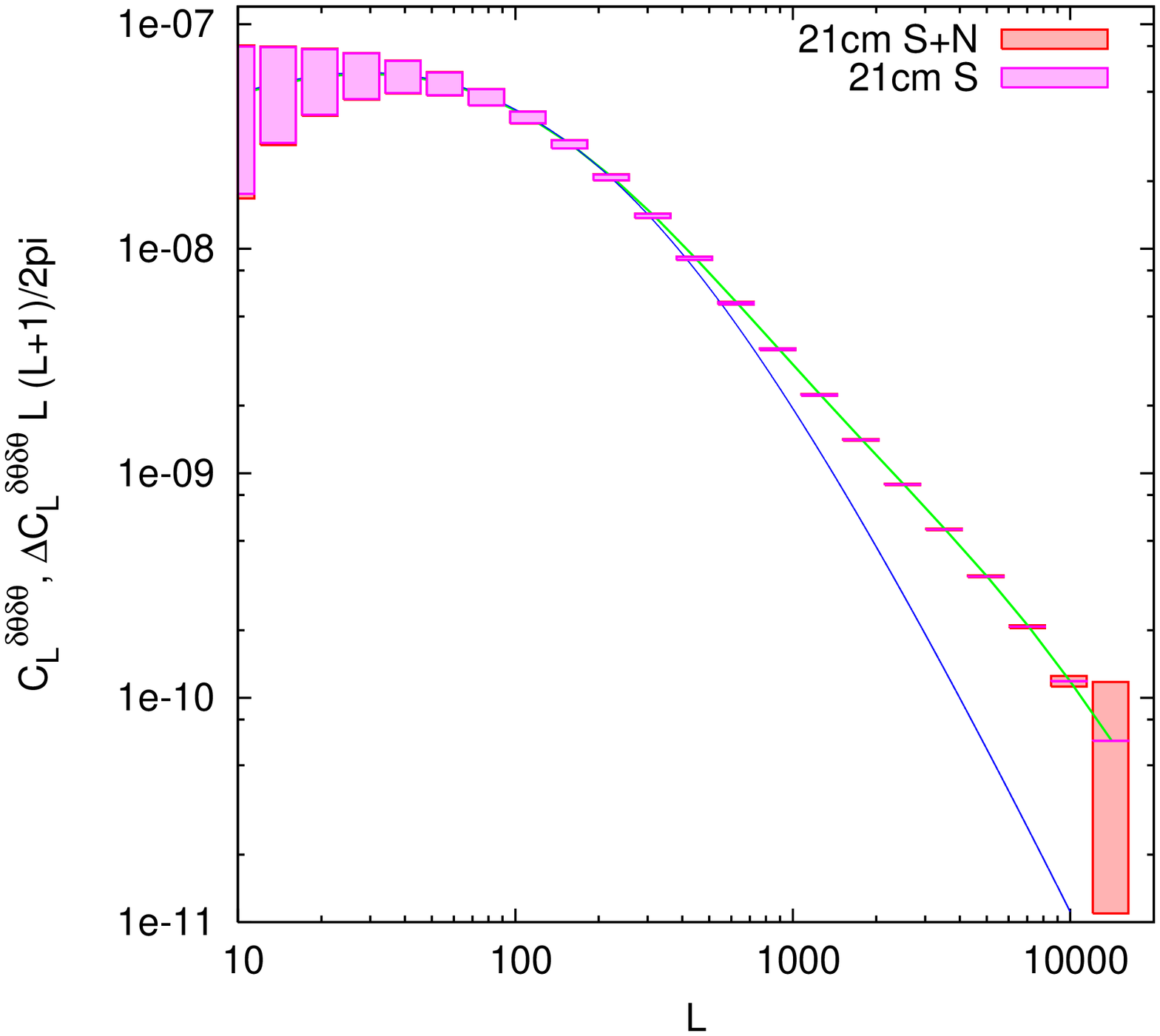}
\ec
\caption{Displacement field power spectra and sample variance (`S') and noise (`N') errors. An SKA like experiment might be
  suited to probe a large dynamical range in the displacement field. The  displacement field angular power spectrum with errorbars on the left is for our
  reference CMB experiment at
  $f_{\rm sky}=0.8$ and the curves on the right are for 21 cm redshift
  6-12 
  lensing reconstruction using the "MWA50k" (same collecting area
  but 10 times higher survey speed than SKA) with $f_{\rm sky}=0.08$. }
\label{fig:deltacl}
\end{figure*}

A comparison between the polarized reference CMB experiment and an
experiment with SKA's sensitivity is shown in Figure 
\ref{fig:deltacl} where we assumed a sky
coverage of $0.8$ for the CMB experiment and a smaller field of view
$0.08$ for the 21 cm experiment. We plot the sample variance error (`S') and sample variance plus noise (`S+N'). Polarized CMB experiments suffer from 
foreground contamination \citep{Page:2006hz}, so our reference experiment should be an idealized limit.
The sky coverage of the 21cm experiment might be achievable within one
year of observation with an MWA 
type experiment that has the same collecting area as SKA but a ten
times higher survey speed. 
 Using fluctuations in the
21 cm background, we should be able to measure the deflection
power spectrum to much higher multipoles, $l > 10000$ for this type of
experiment, than what will ultimately be possible using the CMB
quadratic estimator technique, even if the latter observes on a much
larger part of the sky (notice 
also that 21 cm experiments should be able to observe separate fields
in consecutive seasons).
The requirement of a combination of large collecting area and 
high survey speed makes this an ambitious project.

Finally we would like to estimate the effect foregrounds will have on the
estimation of the lensing field suggested here. As
 pointed out by \citet{Madau:1996cs,Zaldarriaga:2003du}, fluctuations in the gas at high
 redshift can be seperated against the much brighter
 foregrounds (the main source of confusion being galactic and extra-galactic synchrotron), because the former vary rapidly, the latter slowly in
 frequency. This can be done by fitting various smooth functions to
 the signal. \citet{Morales:2005qk}, \citet{Wang:2005zj}, and \citet{McQuinn:2005hk}
 suggest quadratic or cubic polynomials, but also more
 complicated functions such as Chebychew polynomials have been
 proposed. \citet{McQuinn:2005hk} show that this will practically make the
 first few wavevector modes in the line of sight direction unusable. The exact
 number of modes that can be used depends of course on the nature of
 foregrounds, the bandwidth, and the technique of fitting.

 On the positive side once a model for foregrounds is given, we can implement this
 pretty straightforwardly 
 within our formalism by discarding the first few modes $k_\|$ (depending on the order of fitting polynomial) in the
 hierarchy of $C_l$'s, compare  
Figure \ref{fig:clnoise}. It is to be expected that a large number of
 modes still will contribute to the final noise level. This is 
quantified in Figure \ref{fig:NL-hierarchy} where
the sum $N_L^{-1}=1/\sum_{k_{j>j_\min}} N_{L,k_j \mathcal{L}/(2\pi)}$
 is plotted for the two cases $j_\min=3,10$ 
 meaning that the first 3/10 modes have been discarded in the
 analysis. The resulting noise levels (short 
dashed lines) are somewhat higher than the solid line that was
 obtained from assuming no foregrounds.

\section{Discussion}
\label{sec:disc}

In this paper we have extended the quadratic estimator formalism
to use a three dimensional signal as lensing background. 21 cm
fluctuations from neutral hydrogen prior and during the epoch of
reionization contain an enormous amount of data points. The
correlations induced by lensing into this signal can be used to probe
the intervening matter fluctuations either on a individual object
basis, or statistically to for example probe dark energy models.
To describe fluctuations in the neutral fraction, we used an analytic
model for the morphology of HII regions to demonstrate the
applicability of our method to this regime. 

Our estimator should be complete as long as non linearities in the
signal (e.g. due to the bubbles) are small. The bulk of the
information we use is coming from the neutral phase. 

The first generation of experiments is likely not going to be able to
image the angular fluctuations in 21 cm needed to measure the
lensing effect. However we arrive at good constraints by employing current
specifications of the SKA with a flat antenna distribution. 

In comparison with the CMB, 21 cm fluctuations have a rather
featureless scale invariant unlensed spectrum which leads to a smaller
lensing effect. This can be compensated for by using multiple 
redshift information when probing each individual lens surface. The CMB
quadratic estimator sensitivity can actually be beaten this way, for
example with an SKA type experiment.  

Another possibility would be to combine 21 cm lensing reconstruction
with that from other observables, such as the CMB. Similar to combining
the latter with galaxy shear surveys, this will improve the
constraints on the mass/energy budget or geometry of the universe
significantly. 

If ambitions in the community of observational and theoretical
cosmologists increase in the years to come, and third generation 
experiments will be scheduled, a new prospect would be the
measurement of polarized 21 cm emission. \citet{Babich:2005sb} find that
Thomson scattering of the quadrupole produced 
by the reionized universe produces the largest effect and similar to
the gains by using CMB polarization, this could make the method
suggested in this paper, the three dimensional form of the quadratic
estimator, even more promising.

Larger collecting areas and longer observation times also promise to
allow 21 cm lensing reconstruction to map out the gravitational
potential of individual galaxy clusters. We will attempt to address
this topic in 
a future work.

\acknowledgments
O.Z. thanks Miguel Morales and Adam Lidz for useful conversations. The
authors are supported by the David and Lucile Packard
Foundation, the Alfred P. Sloan Foundation, and NASA grants
AST-0506556 and NNG05GJ40G.

\begin{appendix}

\section{Quadratic estimator applied to a three dimensional observable}

We want to observe the three dimensional field $I(\theta, z)$.
We assume weak lensing
\beq
\il (\theta,z) = \iul (\theta,z)+\dtheta \cdot \nabla_\theta
\iul(\theta,z) + ...
\eeq
where $\il(\theta,z)$ is the lensed, $\iul(\theta,z)$ the unlensed
field. The Fourier transform of this expression is
\begin{eqnarray}
\il(\l,k) &=& \iul+\int \frac{d^2l'}{(2\pi)^2}(\delta\theta
(\l-\l')\cdot i\l)\iul(\l',k) \\
&=& \iul(\l, k) - \int {d^2l'} \iul(\l',k) \phi(\l-\l')(\l-\l')\cdot
\l'
\end{eqnarray}
where we have used that $\dtheta(\L)=i\L \phi(\L)$.
We are looking for a quadratic estimator $\Phi(\L)$ for $\phi(\L)$,
i.e. of the form
\beq
\Phi (\L) = \int \d2l \int \dko \int \dkt F(\l,k_1,k_2,\L)
I(\l,k_1) I(\L-\l,k_2)
\label{eq:quadest}
\eeq
(notice that $F(\l,k_1,k_2,\L)=F(\l,k_2,k_1,\L)$). Because $\delta
\Phi(\L)=\delta \Phi^*(-\L)$ it can also be shown that
\beq
F(\l,k_1,k_2,\L) = F^*(-\l,-k_1,-k_2,-\L)
\eeq
We want to find $F$ such that it minimizes the variance of $\Phi(\L)$
under the condition that its ensemble average recovers the lensing field,
$\langle \Phi(\L)\rangle_\iul = \phi(\L)$. This becomes (to first order in $\phi$)
\begin{eqnarray}
\langle \Phi(\L)\rangle_\iul &=&\int \d2l \int \dkt \int \dkt F(\l,k_1,k_2,\L)
\left[-\langle \iul (\l,k_1) \int \d2l' \iul(\l',k_2)
  \phi(\L-\l-\l') \times \right. \nonumber \\
& & \left. \times (\L-\l-\l')\cdot \l'\rangle  - \langle 
  \iul(\L-\l,k_2) \int \d2l'
  \iul(\l',k_1)\phi(\l-\l')(\l-\l')\cdot \l'\rangle \right]
\nonumber \\
&=& - \int \d2l \int \dko \int \dkt F(\l,k_1,k_2,\L) \left[\int \d2l'
  (2\pi)^2 \delta(\l+\l')(2\pi) \delta(k_1+k_2) \pul_{l,k} \times \right. \nonumber \\
& & \left. 
  \times \phi(\L-\l-\l')(\L-\l-\l') \cdot \l' + \int \d2l'
  (2\pi)^2 \delta(\L-\l+\l')(2\pi) \delta(k_1+k_2) \pul_{L-l,k}
  \phi(\l-\l')(\l-\l') \cdot \l' \right]  \nonumber\\
&=& -\int \d2l \int \dko \int \dkt F(\l,k_1,k_2,\L) (2\pi)
\delta(k_1+k_2) 
\left[\pul_{l,k} \phi(\L) \L\cdot(-\l) + \pul_{L-l,k} \phi(\L)
  \L\cdot(-(\L-\l))\right] \nonumber\\
&=& \int \d2l \int \dko \int \dkt  F(\l,k_1,k_2,\L)  (2\pi)
\delta(k_1+k_2) \left[\pul_{l,k} \phi(\L)\L\cdot\l + \pul_{L-l,k}  \phi(\L)\L\cdot(\L-\l)\right] \, ,
\end{eqnarray}
where e.g. $\pul_{l,k}$ is the power in a mode with angular component $l$ and radial component $k$.
With the requirement that $\langle \Phi(\L)\rangle_\iul=\phi(\L)$ this
leads to the normalization condition
\beq
\int \d2l \int \dko \int \dkt F(\l,k_1,k_2,\L) (2\pi) \delta^D(k_1+k_2)
\left[\pul_{l,k}\L\cdot\l + \pul_{L-l,k} \L\cdot (\L-\l)\right] = 1
\label{eq:normcond}
\eeq

The next step is to minimize the variance 
\begin{eqnarray}
\langle ||\Phi(\L)||^2\rangle_\il &=& \int \d2l\int \dko \int \dkt \int
\d2l'\int \dko' \int \dkt' F(\l,k_1,k_2,\L) F^*(\l',k_1',k_2',\L')
\times \nonumber \\
& & \hspace{5.8cm}\times \langle \il(\l,k_1) \il(\L-\l,k_2) \il(\l',k_1') \il(\L-\l',k_2') \rangle \nonumber\\  
&=& \int \int \int \int \int \int F(\l,k_1,k_2,\L) F^*(\l',k_1',k_2',\L')
(2\pi)^2\delta(\l-\l') (2\pi) \delta(k_1-k_1') \ptot_{l,k_1} \times \nonumber\\
& & \hspace{5.8cm}\times (2\pi)^2
\delta(\l'-\l) (2\pi) \delta(k_2-k_2') \ptot_{L-l, k_2} \nonumber\\
& & + \int \int \int \int \int \int F(\l,k_1,k_2,\L) F^*(\l',k_1',k_2',\L') 
(2\pi)^2\delta(\l-\L+\l') (2\pi) \delta(k_1-k_2') \ptot_{l,k_1} \times \nonumber\\
& & \hspace{5.8cm}\times (2\pi)^2
\delta(\L-\l-\l') (2\pi) \delta(k_2-k_1') \ptot_{L-l, k_2} \nonumber \\
&=& \int \d2l \int \dko \int \dkt (2\pi)^2 \delta(0) F(\l,k_1,k_2,\L)
F^*(\l',k_1',k_2',\L') \ptot_{l,k_1}\ptot_{L-l,k_2} \nonumber \\
& & + \int \d2l \int \dko \int \dkt (2\pi)^2 \delta(0) F(\l,k_1,k_2,\L)
F^*(\L-\l',k_2',k_1',\L') \ptot_{l,k_1}\ptot_{L-l,k_2}
\end{eqnarray}
but from \ref{eq:quadest} we see with the substitution $\L-\l
\rightarrow \l$ that $F^*(\L-\l,k_2,k_1,\L) = F^*(\l,k_1,k_2,\L)$
hence
\beq
\langle ||\Phi(\L)||^2\rangle = 2 (2\pi)^2 \delta(0) \int \d2l \int \dko \int
\dkt F(\l,k_1,k_2,\L) F^*(\l,k_1,k_2,\L) \ptot_{l,k_1}\ptot_{l,k_2} \, .
\eeq
Both real and imaginary part of $||F||^2=F_R^2+F_I^2$ contribute
to this variance, however the condition for the
minimization will only pick out the real part. The solution is found
by minimizing the function
\beq
\langle ||\Phi(\L)||^2\rangle- A_R \times (\, {\rm Equation} \, \ref{eq:normcond})
\eeq
with respect to $F$, where $A_R$ is a Lagrangian multiplier. In steps,
\beq
\frac{\partial (\, {\rm Eq.}\, \ref{eq:normcond})}{\partial F(\l,k_1,k_2,\L)} = A_R \d2l \dko \dkt
(2\pi) \delta^D(k_1+k_2) [\pul_{l,k} \L\cdot \l + \pul_{L-l,k} \L\cdot (\L-\l)]
\eeq
and
\beq
\frac{\partial \langle ||\Phi(\L)||^2 \rangle}{\partial F(\l,k_1,k_2,\L)} =
2 (2\pi)^2 \delta(0) \d2l \dko \dkt 2 F_R(\l,k_1,k_2,\L)
\ptot_{l,k_1}\ptot_{L-l,k_2}
\eeq
so
\beq
F_R(\l,k_1,k_2,\L) = A_R (2\pi) \delta^D(k_1+k_2) \frac{[\pul_{l,k} \L\cdot \l +
    \pul_{L-l,k} \L\cdot (\L-\l)]}{\ptot_{l,k_1}\ptot_{L-l,k_2}}
\label{eq:expforf}
\eeq
and by inserting this into the normalization condition
\ref{eq:normcond} we get that
\beq
A_R = \frac{1}{\sum_k \int \d2l \frac{[\pul_{l,k} \L\cdot \l +
    \pul_{L-l,k} \L\cdot (\L-\l)]^2}{\ptot_{l,k_1}\ptot_{L-l,k_2}}} \, .
\eeq
By using Equation \ref{eq:expforf} we find that the variance becomes
\begin{eqnarray}
\langle ||\Phi(L)||^2\rangle &=& 2 (2\pi)^2 \delta(0) \int \d2l \int \dko \int
\dkt \ptot_{l,k_1} \ptot_{L-l, k_2} F_R^2 \\
 &=& 2 (2\pi)^2 \delta(0) A_R^2 \int \d2l \int \dko \int \dkt
[(2\pi) \delta(k_1+k_2)]^2 \frac{[\pul_{l,k}\,  \L\cdot \l +
    \pul_{L-l,k}\, \L\cdot (\L-\l)]^2}{\ptot_{l,k} \, \ptot_{L-l,k}} \nonumber\\
& & 
\end{eqnarray}
Using $(2\pi)\delta^D(0)=\frac{2\pi}{dk}$ and $\int
 \frac{dk}{2\pi}\frac{2\pi}{dk} \rightarrow \sum_k$, this becomes
\begin{eqnarray}
\langle ||\Phi(L)||^2\rangle &=& 2 (2\pi)^2 \delta(0) A_R^2 \sum_k \int \d2l 
\frac{[\pul_{l,k}\,  \L\cdot \l + \pul_{L-l,k}\,  \L\cdot
    (\L-\l)]^2}{\ptot_{l,k}\, \ptot_{L-l,k}} \nonumber \\
&=& 2 (2\pi)^2 \delta(0) A_R^2 \frac{2}{A_R} \nonumber \\
&=& (2\pi)^2 \delta(0) \frac{1}{\sum_k \int \frac{[\pul_{l,k} \L\cdot \l +
    \pul_{L-l,k} \L\cdot (\L-\l)]^2}{2 \, \ptot_{l,k}\ptot_{L-l,k}}}
\end{eqnarray}

With the definition of the noise power spectrum $N_L^\Phi$, 
\begin{eqnarray}
\langle \Phi(\L) \Phi^*(\L') \rangle &=& (2\pi)^2 \delta^D(\L-\L') N_L^\Phi \\
\, {\rm or  } \, \langle  ||\Phi(L)||^2\rangle &=& (2\pi)^2 \delta^D(0) N_L^\Phi
\end{eqnarray}
it follows that
\beq
N_L^\Phi = \frac{1}{\sum_k \int \d2l \frac{[\pul_{l,k}\, \L\cdot \l +
    \pul_{L-l,k}\, \L\cdot (\L-\l)]^2}{2 \ptot_{l,k}\, \ptot_{L-l,k}}} 
\eeq
and since the variances of estimator $D$ of the displacement $\langle
D\rangle_{\rm ens.}=\dtheta$ and $\Phi$ are just related by $\langle ||D(\L)||^2\rangle =
L^2\langle ||\Phi(\L)||^2\rangle$ we obtain
\begin{eqnarray}
N_L^{D} &=& \frac{1}{\sum_k \frac{1}{L^2}\int \d2l \frac{[\pul_{l,k}
    \L\cdot \l + 
    \pul_{L-l,k} \L\cdot (\L-\l)]^2}{2 \ptot_{l,k}\ptot_{L-l,k}}}\nonumber \\ 
      &=& \frac{1}{\sum_k N_{L,k}^{-1}}
\label{eq:noisesum}
\end{eqnarray}
so we have justified that we can simply sum over seperate k modes to
    arrive at our final lensing noise.

\end{appendix}

\end{document}